# Nonautonomous Dynamics of Acute Cell Injury

Donald J. DeGracia,[1,*] Doaa Taha,[2] Fika Tri Anggraini,[1] Zhi-Feng Huang[2]

[1]*Department of Physiology, Wayne State University, Detroit, Michigan 48201, USA*
[2]*Department of Physics and Astronomy, Wayne State University, Detroit, Michigan 48201, USA*

[*]ddegraci@med.wayne.edu

**ABSTRACT**

Medical conditions due to acute cell injury, such as stroke and heart attack, are of tremendous impact and have attracted huge amounts of research effort. The biomedical research that seeks cures for these conditions has been dominated by a qualitative, inductive mindset. Although the inductive approach has not been effective in developing medical treatments, it has amassed enough information to allow construction of quantitative, deductive models of acute cell injury. In this work we develop a modeling approach by extending an autonomous nonlinear dynamic theory of acute cell injury that offered new ways to conceptualize cell injury but possessed limitations that decrease its effectiveness. Here we study the global dynamics of the cell injury theory using a nonautonomous formulation. Different from the standard scenario in nonlinear dynamics that is determined by the steady state and fixed points of the model equations, in this nonautonomous model with a trivial fixed point, the system property is dominated by the transient states and the corresponding dynamic processes. The model gives rise to four qualitative types of dynamical patterns that can be mapped to the behavior of cells after clinical acute injuries. The nonautonomous theory predicts the existence of a latent stress response capacity (LSRC) possessed by injured cells. The LSRC provides a theoretical explanation of how therapies, such as hypothermia, can prevent cell death after lethal injuries. The nonautonomous theory of acute cell injury provides an improved quantitative framework for understanding cell death and recovery and lays a foundation for developing effective therapeutics for acute injury.

## I. INTRODUCTION

Injury is a perturbation to a biological system that disrupts organization or function at one or more levels of the system. There are two broad classes of injury. Acute injuries are those with a clearly identifiable injury mechanism such as ischemia (a large reduction or cessation of blood flow), mechanical force (as in e.g., traumatic brain injury), chemical toxicity, etc. The injury mechanism typically can be applied with a variable intensity, e.g., the duration of ischemia, intensity of mechanical force, concentration of toxin. Chronic injuries such as cancer or Parkinson's disease are distinguished from acute injuries in that the injury mechanism is generally unknown for chronic injuries. The general focus of research on chronic injuries is to identify the injury mechanism itself. For acute injuries, the research focus is to determine how the injury mechanism kills cells. In this work we focus specifically on acute injuries, for which the term "injury" will be used.

Injury to a biological system can affect any of its structural levels, but the ultimate target is the individual cell. If cells die, the tissue, organ, and organism show defects in proportion to the number of dead cells. Injury is studied by comparing the uninjured condition to the injured condition. A list of the biological differences between the uninjured and injured condition is constructed, generally through the work of many laboratories studying the same injury. Since the target of injury is the cell, the bulk of current biomedical research focuses at this level using techniques from biochemistry and molecular biology. Any given biomolecule may increase, decrease, or be unchanged by the injury or it may be modified e.g. by phosphorylation/dephosphorylation, or any number of possible chemical modifications. Chemical species may be found in the injured cells that are absent from the uninjured (e.g. free radical species). Gene expression inevitably changes as a result of injury. The essence of biomedical research is to go, one by one, through the list of differences between the uninjured and injured conditions and attempt to demonstrate which are causal in cell death and which are epiphenomena.

However, it is precisely this approach that has failed to produce effective treatments to halt the cell death for clinical acute injuries. A typical example is stroke, a form of focal ischemia when blood flow to a portion of the brain ceases, resulting in death of that part of brain tissues. Stroke research has uncovered conditions that predispose to stroke: obesity, heart disease, diabetes, and so on, and led to preventative efforts that have decreased strokes incidence [1]. Research has also led to the development of surgery methods and one chemical intervention (tissue plasminogen activator, TPA, a "blood thinner"), that can alleviate stroke injury to varying extents [2]. However, these methods can only be used on < 10% of the approximately 750,000 stroke patients each year in the USA [3]. The remaining > 90% of stroke victims will suffer irreversible brain damage. To date, there have been close to 150 clinical trials [4] to test therapies to halt or slow stroke-induced brain damage in the stroke patients who cannot undergo surgery or TPA treatment. Every single stroke clinical trial has failed [5]. This situation is not unique to stroke but applies to a diverse array of medical conditions: cardiac arrest [6], myocardial ischemia ("heart attack") [7], acute kidney injury [8], traumatic brain injury [9], and so on.

Although the current research approach has produced a wealth of biomedical information at the cellular, subcellular, and biomolecular levels, it is qualitative, describing altered molecular pathways and coupling among different cellular factors. This approach does not account for quantitative parameters, such as the injury intensity, and how any such parameters link to cell recovery or death. It is also an inductive approach, relying exclusively on the specifics of measurements and observations and being limited by the degree of completeness of empirical information. This approach too often conflates a difference detected between the injured and uninjured conditions with the cause of cell death. In addition, when one aspect of injury is manipulated (e.g. by giving a drug that targets that aspect) there is little consideration given to how this would affect all of the other changes on the list of differences between the uninjured and injured conditions. That is, the notion of "all other things being equal" is commonly assumed to hold in biomedical research. However, this assumption is untested, and given the complexity and network-type characteristics of biological systems [10], unlikely to be true.

These weaknesses can account, at least in part, for the failures to find effective treatments for important medical conditions listed above. An influential metareview that critiqued stroke therapies developed on the qualitative basis described above concluded: "[That]...*no particular drug mechanism distinguished itself on the basis of superior efficacy...might suggest that our conception of stroke needs reformulation*" [5]. To fully understand the importance of this quote, we briefly review salient facts about acute injuries.

**A. Salient facts about acute injuries**

*1. Outcome is binary*

Outcome refers to the disposition of individual cells after injury: individual cells either recover or die after injury. There is not a third outcome. If a cell survives being injured, it does not permanently transform into a phenotype different from what it was before the injury. The ideal goal of medically treating an injury is to foster the injured cells to resume their preinjury phenotype so that the organ can function as it did before the injury. If injured cells transformed into a phenotype different from the preinjury phenotype, then this medical goal would not be possible. There are two points of possible confusion with the assertion that outcome is binary: (1) surviving cells in an organ adapt to the loss of cells that died from the injury, and (2) cells undergo transient changes after injury. We briefly discuss each in turn.

*2. Adaptations*

An example of postinjury adaptation is hypertrophy following cardiac ischemia [11]. Ischemia of the heart causes contractile cardiomyocyte cells to die. A normal heart generates specific forces with each beat. After cardiac ischemia, these forces are altered and diminished due to death of cardiomyocytes. The surviving cardiomyocytes will grow in size to attempt to compensate for the lost cells, a condition called hypertrophy. The surviving cells do not assume this new phenotype as a result of being injured, but as a result of the loss of cells that formerly contributed contractile force in the heart. That is, the stimulus for this change is not the injury intensity, but the altered mechanical forces in the heart following the death of other cardiomyocytes. If the cardiomyocytes that died could have been prevented from dying, there would have been no stimulus to trigger the hypertrophy adaptation. Such adaptive situations are secondary to the primary injury, and the models of cell injury studied here are not intended to account for such cases.

*3. Transient dynamics*

It is well-established that injured cells show transient changes, given that the system of interest is intrinsically nonequilibrium. Our models are designed to describe these transient post-injury dynamics. A most important transient change, the "preconditioned" state, will be discussed below. Usually in a standard nonlinear dynamical analysis, the final states of the system are of primary interest, and the transient dynamics towards those states are of secondary importance. In the study of acute cell injury, it is the opposite case. The final states are, in essence, trivial: the system is binary and either reverts to its preinjury state or dies. It is the trajectory followed by the cell after injury that is of central interest, and we aim to account for the behavior of the cell in the postinjury period because that is the time when medical intervention is possible. If the transient trajectory of the system is understood, then it is possible in principle to perturb that trajectory to achieve some desired aim. For example, if the system is on a transient trajectory towards the death outcome, knowledge of that trajectory may allow an intervention to change it from the death outcome to the survival outcome. Thus, a novel feature of our study here is an explicit focus on the transient post-injury dynamics, and not on the final states of the system.

*4. Quantitative parameters determine outcome.*

Whether a cell recovers or dies is a function of the injury intensity, $I$. To concretely visualize a range of injury intensity, consider a series of the same cell type where each member is injured at increasing intensity by some generic injury mechanism. Cell 1 is injured with $I_1$, cell 2 injured with $I_2$, and so on with cell $N$ injured by $I_N$. For intensity below a lower threshold $I_{min}$, which is a mere increment over zero, the injury may have no effect at all on the cell. At a very high $I = I_{max}$, the injury will be so intense that the cell is instantly destroyed. It is obvious that as $I$ increases between $I_{min}$ and $I_{max}$, the amount of damage to the cell will also increase, and there will be a continuum of the cell's response dynamics along the range $I_{min} < I < I_{max}$, which henceforth will be called the $I$-range. A major goal of the cell injury models is to quantitatively describe the injury dynamics across the $I$-range, as will be detailed below.

*5. Injury produces many changes in the injured cell.*

What is the physical response of a cell to being injured? The previous section insinuates that the answer to this question depends on the parameter $I$. While this is true, it is nonetheless possible to abstract and generalize the possible responses displayed by an injured cell. There are only two possible categories of response: (1) the cell is passively damaged by the injury, and (2) the cell actively responds to being injured.

*6. Cell damage*

By "damage" we mean disruption of the structural and/or functional organization of a cell. Structural damage includes breakdown of cell components or loss of structural organization such as breakdown in the various compartments of a cell. Damage can also be functional, e.g. a disruption in temporal or enzymatic activities. In practice, both structural and functional disorganization occurs after injury. Empirically, there is a long list of possible forms of cellular damage. However, because cells are finite, the list of how they can be damaged is also finite. Any specific injury mechanism will draw from this ensemble of possible forms of damage, and it is expected that different injures will produce overlapping sets of damage products. The common feature of any form of damage produced by any injury mechanism is that the damage is a *passive* by-product of injury to the cell.

*7. Cell stress responses*

However, cells are not passive bystanders in the face of injury. They actively seek to retain their integrity if it is disrupted. In biology, this is called homeostasis and consists of a long list of possible mechanisms a cell can exert to overcome disruption. It is a spectrum between a homeostatic perturbation and frank injury, again implying a continuum of injury intensities.

It has become clear over the past decades that cells respond to injury by stress reprogramming [12]. When a cell is injured, it possesses molecular mechanisms to detect the passive damage products. These mechanisms alter the genetic programming of the cell by changing the set of mRNAs produced, and therefore the corresponding proteins inside the cell. Such stress response reprogramming mechanisms are the physical basis for the transient responses exhibited by injured cells. The function of stress responses is to ameliorate the damage caused by injury by inhibiting damaging chemical reactions, rebuilding damaged parts of the cell, clearing out damage products, etc. [13].

We give one example of stress reprogramming. The heat shock response (HSR) was discovered in the early 1960s as a cellular response after heating cells to potentially lethal temperatures (e.g. from 37º C to 42º C) [14]. It is now known a large variety of injuries activate the HSR: heavy metal toxicity, oxygen deprivation, glucose deprivation, ethanol exposure, etc. [15]. The HSR is an ancient genetic program, highly conserved from yeast to humans, and even present in bacteria [16]. The HSR is a reprogramming event where the injured cell ceases translating its native proteins for a transient period, which for mammalian cells is on order of 12-24 hours after the injury. During this period the cell instead transcribes and translates almost exclusively a set of 8-12 proteins called heat shock proteins (HSPs) [17]. HSPs belong to a class of enzymes known as chaperones, which are proteins that assist other proteins to fold into their proper 3-dimensional structure. The HSR is triggered by abnormally high concentrations of denatured proteins in the cell. The various injury mechanisms listed above, although appearing superficially diverse, all cause denaturation of the cell's proteins (each by different specific mechanisms), and thereby a diverse range of injury mechanisms trigger the same stress response across almost all living organisms.

A couple dozen stress responses have been discovered so far. Along with the HSR, others are the unfolded protein response [18], the anti-oxidant response [19], the mitochondrial stress response [20], several distinct DNA repair programs [21,22], and so on. They share the feature that diverse insults activate them across diverse species. Since all cells evolved from a common ancestor, and because genomes are finite, it is expected that there will be overlap in the response of all cells to a variety of injury mechanisms. There will be species-specific variations, but these will be smaller when the evolutionary difference between species is smaller.

A general feature shared by all stress responses is that they are mediated by proteins produced by the cell. There is only a finite capacity of these proteins and they can be saturated if the concentration of the damage product greatly exceeds the concentration of stress response proteins. Our model below accounts for the saturation of stress responses.

To summarize, cells are made of the same physical substances (nucleic acids, proteins, lipids, sugars, etc.) with the same general plan of organization. Cells utilize evolutionarily-related homeostatic mechanisms. Thus, diverse cell types injured by different forms of injury exhibit overlapping sets of many passive damage products and active stress responses.

**B. Therapy After Acute Cell Injury**

In biomedical research, no matter what specific injury is studied, the goal is the same: to develop a successful therapy for that injury. In this context, therapy means to take an injured cell for which it is known it will die following the injury and perform some intervention to prevent the cell from dying. This is called a "protective" therapy. This definition of protective therapy has several antecedents. First, only at very high injury intensity does cell death occur during the injury. At lower injury intensities, substantial amounts of time can pass between the end of the injury and the occurrence of cell death (see the example of stoke discussed below). Because there is time between injury and cell death, it is possible to carry out a therapeutic intervention to halt the cell death, and any such intervention is a protective therapy.

It is well-established that the rate cells die following acute injury varies, and the time between injury and cell death depends on the injury intensity [23]. For example, if stroke duration is very short (e.g., on order of 5 minutes, a "transient ischemic attack"), there will be no cell death [24]. If stroke duration is very long (e.g., 8 hours), the cells die during the ischemia [25]. If ischemia is moderate (e.g., 1 hour), then it may take days, weeks, or even months after the end of the ischemia for brain cells to die [26]. Rapid cell death after injury is called necrosis. In stroke research, the slower death process is called delayed neuronal death. It is not a binary distinction between necrosis and delayed death because the delayed death occurs over variable durations and is therefore more like a continuum.

In the case of stroke research, the therapies mentioned above, surgery and TPA, serve to shorten the ischemia duration and thereby decrease the injury intensity. They do not satisfy the definition of protective therapies because, by decreasing injury intensity, the cells are prevented from being exposed to a lethal injury intensity [3]. However, lowing injury intensity is only effective if the patient presents to a physician within the first 4 hours after stroke onset, which happens in only roughly 10% of stroke cases, while the typical stroke patients present within the first 24 hours after stroke onset [27]. In this typical case, the brain cells have been exposed to lethal injury intensities. The patient presents to the physicians with a region of dead, necrotic brain tissue called a "core". Neurons do not divide and thus the core is unsalvageable. In addition, over the ensuing days, weeks, and months, it is observed that the size of the core region grows in the brain of the patient, which is the manifestation of delayed death. This is a significant clinical example where it is known beforehand that the cells will die by delayed death and there is ample time to intervene to stop the cell death. The goal of stroke neuroprotection is to prevent the growth of the core over the ensuring days and weeks after a stroke [28]. It is this effort, to prevent delayed neuron death and the growth of the core, that has shown a 100% failure rate in clinical trials [4,5].

The therapies tested in clinical stroke trials were drug treatments. The drugs chosen were based on the list of biological differences between normal brain and brains subjected to stroke in experimental animals. The drugs targeted a diverse variety of qualitatively different cellular mechanisms. To name a few of the specific mechanisms targeted [4,5]: free radical scavengers; drugs that decreased swelling of the brain; growth factors that stimulate protective patterns of gene expression; drugs that alter the flow of $Ca^{2+}$ ions in the tissue; etc.. The list of tested drugs is long. But as the quote above stated: "*no particular drug mechanism distinguished itself on the basis of superior efficacy*" [5]. That is, of the specific mechanisms in the list of differences between a normal brain and a stroke brain, none of them stood out as more effective at stopping delayed cell death than the other. When used in clinical human stroke trials, none of them worked to prevent the growth of the core region over time.

However, the differences between stroked and normal brain are empirical facts that have been reproduced across different laboratories. They are not false experimental results. The key point is that stroke induces many forms of cellular damage and many forms of cellular stress responses in the brain cells [29]. The realization that has emerged from the failure of stroke clinical trials is that if the therapy treats only a single form of damage, it is ineffective at preventing cell death. Some physicians and researchers have advocated multiple-drug therapies [30]. However, this suggestion is unrealistic because: (1) specific drugs cannot be given for the hundreds of identified biological changes in the injured tissue (e.g. as reviewed in [29]), and (2) increasing the number of drugs administered simultaneously compounds drug interaction effects, leading to an unpredictable and unmanageable situation. There are, however, two protective therapies that are effective at stopping cell death after injury, including stroke, as we now describe.

**C. Preconditioning and Hypothermia**

Preconditioning and hypothermia have proven effective at preventing cell death across diverse injuries applied to diverse organs and cell types. Preconditioning is a situation where the cell or organ is first exposed to a sublethal injury intensity and then, after some duration, is exposed to a second injury at a lethal injury intensity [24,31]. Prior exposure to the sublethal insult prevents cells from dying after exposure to the lethal insult. The cells become "conditioned" from the first, sublethal insult (the "pre-" insult), hence the name preconditioning. The preconditioning protocols, however, vary across cell types. For example, in brain, a typical protocol is to perform a short, sub-lethal ischemia (2 min), wait two days, and then perform a lethal (10 min) ischemia [32]. The two-day intervening time period is important. If it is shorter or longer than two days, the effect is suboptimal. If the time between the sublethal and lethal insults is one week or greater, the preconditioning effect is lost. Thus, preconditioning is a transient effect. A protocol to precondition heart cells is to perform three 1-min, sublethal ischemias, each separated by 5 mins, followed by a lethal 10-min ischemia on the heart, which prevents the death of cardiomyocytes [33]. For different organs the effect of preconditioning is the same: it prevents cell death after the lethal insult. But the temporal parameters required to optimize the effect are vastly different between brain and heart, indicating that each exhibits different injury dynamics with respect to the same injury, ischemia, in this case. Such protocols are discovered by trial and error and at present there is no effective way to determine beforehand how to optimize such protocols. Our models below seek to provide a deductive framework that ultimately might be used for making such determinations.

Similarly, lowering temperature, i.e., hypothermia, has shown to be highly effective at preventing the death of cells that have experienced an otherwise lethal injury. Hypothermia is already in clinical use for transplantation surgeries [34] and some cardiac surgeries where keeping the organs cold prevents cell death during periods of no blood flow [35,36].

There are ongoing research efforts across many laboratories to understand how preconditioning and hypothermia prevent cell death. If the mechanism of their protection could be understood, it could, in principle, be exploited to stop cell death in other cases. However, this research proceeds via the qualitative list approach described above, which now generates a more complex situation. Not only must a list be constructed for the differences between the uninjured and lethally injured tissues, but also lists of differences between the uninjured and preconditioned (or hypothermic) ones, the preconditioned vs. injured, and so on. The number of list comparisons multiplies and becomes unmanageable. In general, the list-comparison approach has failed to find "the" biological mechanism of either preconditioning or hypothermic protection. Instead, research into the protective effects of these treatments has demonstrated, as in the case of untreated injury, that many cellular and molecular changes can be detected between the uninjured, lethally injured, and the preconditioned (or hypothermic) cases [37-39].

Preconditioning and hypothermia share a vital common feature: they are both forms of injury to cells. Preconditioning is so by definition. Hypothermia, as applied to biological systems, is also a mild form of injury. It is an injury mechanism of variable intensity, where the intensity of insult is the temperature difference. A mild temperature differences prevents cells from dying, but a large temperature difference kills cells.

Thus, we come to a surprising conclusion that the most effective way discovered so far to prevent cell death after a lethal injury is to provide a prior sublethal injury. Similar to the case with stroke, no specific molecular or cellular mechanism has stood out as "the" mechanism that generates the protective effects of either preconditioning or hypothermia. The deductive models presented below, both autonomous and nonautonomous, offer a new dynamical understanding of the protective effects of preconditioning and hypothermia. The autonomous model suggested to us they are perturbations that shift a system on a death trajectory to a survival trajectory. The nonautonomous model suggests that perturbations like preconditioning and hypothermia access a reserve stress response capacity that is likely critical for achieving a survival outcome.

The above summarizes the qualitative findings across many biomedical disciplines, using stroke research as our focal point. The qualitative understanding of how cells respond to injury has advanced enough that it is now possible to abstract and generalize the findings in the service of constructing deductive models that can help organize what is otherwise a bewilderingly complex array of specific biological findings (such as those in Ref. [40] that sought to classify cell death on qualitative bases).

## II. A DEDUCTIVE APPROACH TO CELL INJURY

We begin by noting that we consider the deductive theories developed here to be coarse-grained phenomenological models. They contain over-simplifying idealizations and are based on a series of assumptions that ultimately require empirical verification. Our justification for developing and studying phenomenological models is the fact that the qualitative list-comparison approach described above has failed to generate successful clinical therapies; therefore alternative approaches are required. With this deductive approach we aim to: (1) determine if deductive dynamical models can capture the salient aspects of cell injury discussed above, and (2) determine if a study of the phenomenological models offers new insights into acute cell injury.

The first model is an autonomous system of coupled ordinary differential equations (ODEs) that we have developed before [41]. In the following we provide an overview of the theory, pointing out the novel insights provided by this autonomous model and also highlighting its weaknesses. The main purpose of this paper is to describe the global dynamics of the nonautonomous extension of the theory and show how it overcomes the limits of the autonomous model and provides a fuller understanding of the acute injury dynamics. As will be shown below, this nonautonomous model exhibits different and more complicated dynamic behavior as compared to those known from standard nonlinear dynamics, and the standard fixed-point and bifurcation analyses could not be applied since the system property here is determined by transient dynamic processes. On the other hand, the resulting dynamic patterns can well represent various clinically-relevant cell behaviors after injury, and a new quantity that is important to therapies, i.e., a latent stress response capacity (LSRC), can be identified.

### A. The Autonomous Model

As described in Sec. I A, the key empirical finding at the core of our model is that acute injury simultaneously induces many forms of damage and activates many stress responses in the cell. We formalized these observations via the concepts of **total induced damage, *D*** , and **total induced stress responses, *S***. These may be symbolically expressed as

$$D = \sum_i d_i,$$
$$S = \sum_i s_i, \qquad (1)$$

where $d_i$ are all the individual forms of passive damage induced by the injury of intensity $I$, and $s_i$ are all of the cell's active responses to the injury. No matter what specific forms of damage are produced by injury, the aggregated or overall effects of them give the total amount of damage produced in the cell, and similarly for the induced stress responses. Thus, instead of listing the fine-detailed changes, we introduced a course-graining approach to describe the cell's response to injury. Eq. (1) is a conceptual definition. Converting it to an operational definition is beyond the scope of this paper and is not necessary for the theoretical analysis offered here.

For a given injury intensity, $I$, the individual $d_i$ and $s_i$ will follow their intrinsic time courses, causing $D$ and $S$ to change with time. We idealized application of injury with intensity $I$ to be an instantaneous event at time $t = 0$, after which the changes in $D$ and $S$ constitute the system dynamics. Additionally, by definition, $D$ and $S$ are mutually antagonistic. Damage will blindly act to destroy any aspect of cell function, including stress response mediators. Stress responses exist to eliminate cell damage as described above in Sec. I. To model how the mutual antagonism of $D$ and $S$ change with time, we used a well-known system of coupled nonlinear ODEs that specifies a winner-take-all competition [42,43]. The net rates of change of $D$ and $S$ equal their formation rates minus decay rates, i.e.,

$$\frac{dD}{dt} = v_D \frac{\Theta_D^n}{\Theta_D^n + S^n} - k_D D,$$
$$\frac{dS}{dt} = v_S \frac{\Theta_S^n}{\Theta_S^n + D^n} - k_S S, \quad (2)$$

where the $v$ and $k$ parameters scale the rates of formation and decay, respectively. The threshold $\Theta_D$ is the amount of $D$ that decreases $S$ by 50%, and $\Theta_S$ is the amount of $S$ to decrease $D$ by 50%. The parameter of exponent, $n$, can be taken as a measure of the coupling amongst the individual $d_i$ and $s_i$ of which $D$ and $S$ consist. Note that time $t$ in all the model equations presented in this paper is rescaled and the real unit can be determined only when compared to experimental data which is part of our ongoing work of $D$ and $S$ measurements and will be reported elsewhere.

We briefly justify the use of Eq. (2), similar forms of which have successfully described dynamics of other biological systems where mutual antagonism is a core feature (as described in a general context in Ref. [42]). Huang *et al.* [43] showed that a similar version of Eq. (2) could model transcription factor dynamics governing the binary fate decision of a stem cell, and Tawari *et al.* discuss using such models to describe switching dynamics in bacterial gene networks [44].

Next, we posited that the thresholds $\Theta_D$ and $\Theta_S$ change as a function of injury intensity, $I$. As indicated above, the meaningful $I$-range is $I_{min} < I < I_{max}$. As $I$ increases, there will be an increase in the total amount of damage to the cell, causing the corresponding increase in $\Theta_D$. The larger $\Theta_D$ is, the harder for $S$ to overcome $D$. The idea that stress responses are finite, and hence can saturate, was discussed above. Therefore, $S$ is expected to increase with $I$ to a point and then decrease with increasing $I$ thereafter because the excess damage saturates the finite stress responses. $\Theta_S$ should follow the behavior of $S$. In our modeling we assumed $\Theta_D$ increases exponentially with $I$ and $\Theta_S$ decreases exponentially with $I$, i.e.,

$$\Theta_D = c_D I e^{I\lambda_D},$$
$$\Theta_S = c_S I e^{-I\lambda_S}, \quad (3)$$

with the scaling parameters $c_D$, $c_S$ and the exponential parameters $\lambda_D$ and $\lambda_S$. The parameters ($c_D$, $\lambda_D$) can be taken to give a quantitative representation of the specific form of injury (e.g., ischemia vs mechanical trauma, etc.), and ($c_S$, $\lambda_S$) can be taken to represent a specific cell type (e.g., neuron vs cardiomyocyte, etc.), as discussed before [45]. We stress that the functional form of Eq. (3) is merely for the sake of modeling.

The true relationship between the thresholds and $I$ requires empirical investigation.

The autonomous model of acute cell injury was obtained by substituting Eq. (3) into Eq. (2) and assuming $v_D = v_S = v$ and $k_D = k_S = k$ for simplicity, giving

$$\frac{dD}{dt} = v \frac{(c_D I e^{I\lambda_D})^n}{(c_D I e^{I\lambda_D})^n + S^n} - kD,$$
$$\frac{dS}{dt} = v \frac{(c_S I e^{-I\lambda_S})^n}{(c_S I e^{-I\lambda_S})^n + D^n} - kS. \qquad (4)$$

### B. Solutions of the Autonomous Model

Equation (4) is nonlinear and has no closed form solution, therefore solutions were obtained by standard numerical methods. From Eq. (3), a parameter pair $(c_D, \lambda_D)$ was interpreted to indicate a specific injury mechanism, and a pair $(c_S, \lambda_S)$ interpreted to indicate a specific cell type [45]. The $v$ and $k$ parameters were set to 1 to scale solutions to the unit plane [41]. The value of $n$ was arbitrarily taken as 4, since varying $n$ over $1 \leq n \leq 25$ does not affect the qualitative dynamics [43]. Thus, a parameter input vector consists of $(c_D, \lambda_D, c_S, \lambda_S, n=4, v=1, k=1, I)$. The goal was then to study the system dynamics over the $I$-range.

The relevant $I$ range was determined by calculating the tipping point value of $I$, $I_X$. $I_X$ is the value of $I$ where for $I < I_X$, the cell survives and for $I > I_X$, the cell dies [41]. $I_X$ is calculated by setting $\Theta_D = \Theta_S$ in Eq. (3), which leads to

$$I_X = \frac{\ln(c_S) - \ln(c_D)}{\lambda_D + \lambda_S}. \qquad (5)$$

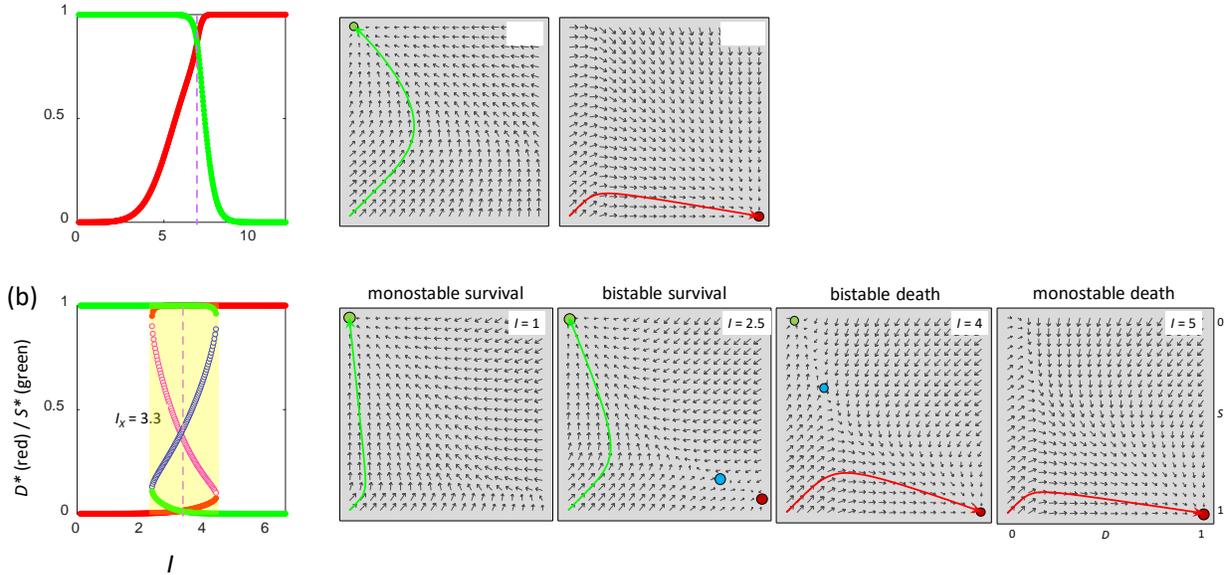

FIG 1. (a) Monostable and (b) bistable bifurcation diagrams obtained from Eq. (4), and the representative phase planes at indicated $I$ values. In the bifurcation diagrams, purple dashed lines indicate $I_X$, and green/red are attractors. In the first panel of (b), blue/red open points are repellers for $S$ and $D$, respectively, and yellow area indicates bistable part of $I$-range. In phase planes, trajectories are from $(D_0, S_0) = (0, 0)$, with survival attractor (green), death attractor (red), and repeller (blue). Scale on $I = 5$ phase plane in (b) applies to all the phase planes. Parameters $(c_D, \lambda_D, c_S, \lambda_S, n)$ are for (a) (0.1, 0.1, 100, 0.9, 4) and for (b) (0.075, 0.1, 2, 0.9, 4), with all other parameters equal to 1.

Bifurcation diagrams of Eq. (4) were constructed with $I$ as the control parameter. This was meant to model application of the injury mechanism ($c_D$, $\lambda_D$) to some cell type ($c_S$, $\lambda_S$) at each $I$ across the $I$-range. The fixed point ($D^*$, $S^*$) at each $I$ determines the outcome of the cell. If $S^* > D^*$, then the final state shows the stress responses dominating over damage, corresponding to the survival outcome. If $D^* > S^*$, then damage dominates over stress responses, and the outcome was death. Many combinations of ($c_D$, $\lambda_D$) and ($c_S$, $\lambda_S$) were studied across the relevant $I$-ranges. A major finding of the qualitative analysis was that Eq. (4) output only 4 qualitative types of bifurcation diagram [41]. The two canonical types are illustrated in Fig. 1. The other two types are variants of those shown and not discussed here.

Figure 1(a) illustrates a monostable bifurcation diagram exhibiting a single fixed point attractor at each $I$. For $I < I_X$, $S^* > D^*$, and the outcome was survival. For $I > I_X$, at each attractor $D^* > S^*$, representing death outcomes. There was one value of $I = I_X$ where $D^* = S^*$, thereby demonstrating $I_X$ as the tipping point between the survival and death outcomes. The other panels in Fig. 1(a) show individual phase planes at $I = 3$, and 10, respectively, indicting the respective trajectories from the initial condition ($D_0$, $S_0$) = (0, 0). The bifurcation diagram shows that monostable systems have two possible "phases": either survival at $I < I_X$ or death at $I > I_X$.

Figure 1(b) illustrates a second type of bifurcation diagram that exhibited bistability. In part of the $I$-range, two attractors, representing both the survival and death outcomes, are present on the phase planes (yellow bar in the middle of first panel). Four phase planes are shown in adjacent panels, with bistable phase planes shown at $I = 2.5$ and 4. Prior to the bistable region of the $I$-range, all outcomes were survival, and after the bistable region, all outcomes were death. For $I < I_X$ in the bistable region, trajectories from initial conditions of (0, 0) or of large enough $S_0$ survived, although those from large enough $D_0$ with small enough $S_0$ would die. For $I > I_X$ in the bistable region, trajectories from (0, 0) or large enough $D_0$ died but those from large enough $S_0$ with small enough $D_0$ survived. Thus, the bistable bifurcation diagram exhibited four "phases": monostable survival, bistable survival, bistable death, and monostable death. $I_X$ was centered in the bistable range, and $D^* = S^*$ occurred as the repeller at $I_X$.

Figure 2 illustrates how the durations of the trajectories from initial condition (0,0) to the attractor state varied with $I$ for the bifurcation diagrams in Fig. 1. The durations were constant at low $I$ and then increased as $I$ approached $I_X$. After $I_X$, there is a continuous decline in the durations until at high enough $I$ the durations became constant again and were the shortest.

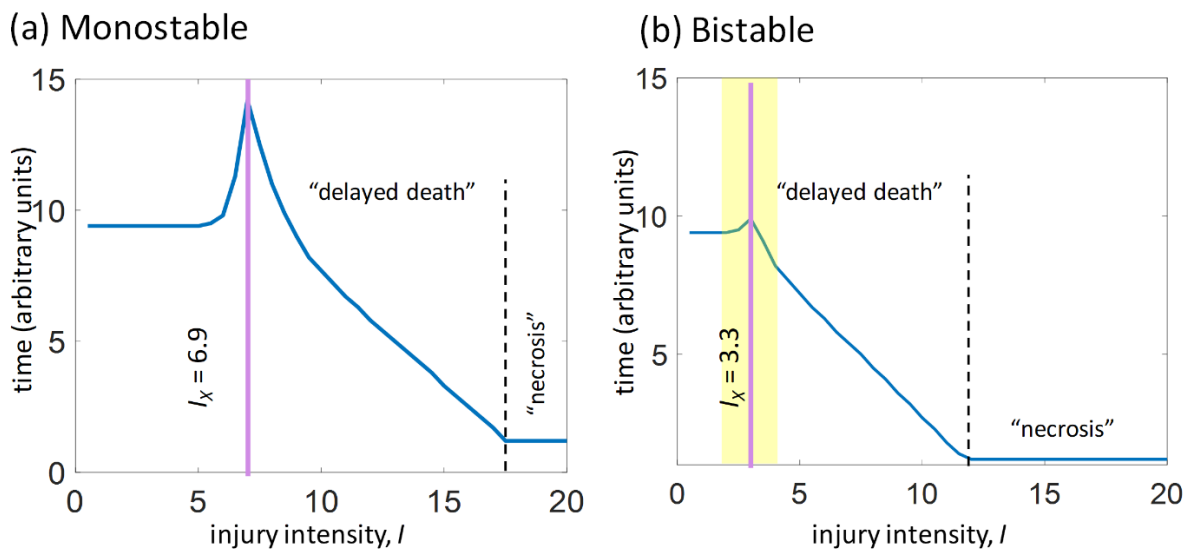

FIG 2. Time to the attractor states for trajectories starting from ($D_0$, $S_0$) = (0, 0) for (a) monostable and (b) bistable systems shown in Fig. 1. Bistable region is indicated by yellow bar in (b). $I_X$ is indicated by purple solid line. "Delayed death" durations are indicated between $I_X$ and the black dashed line that separates them from the rapid "necrosis" duration. Time is expressed in arbitrary units for this and all figures containing time plots.

### C. Insights from the Autonomous Model

Two key insights emerged from studying Eq. (4): (1) the interpretation of bistable solutions in the context of cell injury, and (2) the variation of time course durations with $I$. The occurrence of bistability was the most important finding because it suggested an entirely new mechanism of therapy. Therapy was defined above as performing some intervention on an injured cell that is fated to die and preventing cell death. The deeper question is: how is this physically possible? In the qualitative biomedical approach, therapy is explained by specific biological mechanisms. For example, after injury there may be a large increase in free radical species that are presumed to cause death by destroying cell components. Therefore, if a drug is given to halt free radicals (the therapy), then cell death should be prevented. However, this approach has failed to produce success in clinical trials for the reasons discussed above.

The results shown in Fig. 1(b) provide the completely new insight that therapy is a consequence of bistable injury dynamics. When $I > I_X$, these systems will always die from initial condition $(0, 0)$. On the other hand, if the system is in the bistable range of injury intensity $I$, both attractor states are possible outcomes of the system. This result allowed us to envision that therapy could be thought of as a perturbation on the trajectory from $(0, 0)$ to the death attractor that diverts the system to the survival attractor [46]. This suggested to us a dynamical conception of therapy where the biological factors were not causative in outcome, but instead instantiate the dynamics. Whether a system is amenable to therapeutic treatment would depend on the dynamics. In the scope of the autonomous model described above, therapy is only possible when the dynamics are bistable. We have discussed the application of this concept in other publications and refer the interested readers to them [41,46,47]. However, although the autonomous model allow envisioning these insights, it cannot implement this concept of therapy, i.e., diverting a system from a pro-death to a pro-survival trajectory. Developing a theory that allows implementation of this dynamical concept of therapy is therefore the main motivation for the current work.

The second insight obtained from the autonomous model is that the results in Fig. 2 resemble the empirical fact that the time to death after injury varies with injury intensity. There is a range of $I > I_X$ that shows relatively longer but variable durations, resembling the delayed death phenotype (labeled "delayed death" in Fig. 2). At high $I > I_X$, the durations become constant for all subsequent $I$ and are relatively rapid, resembling the necrosis phenotype (labeled "necrosis" in Fig. 2). The word "resemble" is used because the result does not recapitulate the empirical facts. However, these resemblances were important clues in the development of the nonautonomous model, which does recapitulate the empirical facts as described below.

### D. Weaknesses of the Autonomous Model

The study of durations shown in Fig. 2 reveals a weakness of Eq. (4): It does not describe the full time course following injury. When the cell is uninjured, there is no damage or stress responses, i.e., $D = S = 0$. To capture the full time course after injury, the model must allow the system to return to $(0, 0)$. This is true regardless of whether the system survives or dies. For survival, when the cell recovers from injury both $D$ and $S$ return to 0 (the uninjured state). In the case of death, all cell variables go to 0 because the cell disintegrates. However, Eq. (4) puts the injured cell to a final state $(D^*, S^*) \neq (0, 0)$. Thus, having trajectories begin and end at $(0, 0)$ was an important technical consideration in developing the nonautonomous model.

The second weakness of Eq. (4) lies in the interpretation of the initial conditions. The meaning of $D_0 > 0$ is that there is pre-existing damage in the cell at the instant injury of intensity $I$ is applied. In this case, we would expect the pre-damaged cell ($D_0 > 0$) to be weaker compared to a cell with no pre-existing damage ($D_0 = 0$). However, in the monostable phase planes for $I = 3$ in Fig. 1(a), and $I = 1$ in Fig. 1(b), for all $D_0 > 0$ the outcome is always survival. This is an unrealistic result. If a cell had severe enough damage before application of even a low intensity injury, it is expected that there would exist initial conditions from which the cell would die, even if $I < I_X$. Therefore, the model would be improved if outcome was not only a function of injury intensity, $I$, but also of the initial conditions $(D_0, S_0)$. We will show below this feature emerges automatically from the nonautonomous model.

Figure 2 is one way to represent the family of time courses from $(D_0, S_0) = (0, 0)$ across the $I$-range. This family of time courses has a special significance by representing the "natural" response of a cell to an injury across the full range of injury intensities. As indicated above, nonzero initial conditions imply either pre-existing damage ($D_0 > 0$) or pre-activated stress responses ($S_0 > 0$) in the cell. By what physical means can initial conditions be altered? While there are many specific means, they can be generalized as forms of pre-treatment of the cell before application of the injury. A canonical pretreatment is preconditioning, which, as described above, prevents cell death after application of a subsequent injury that would otherwise be lethal. Thus, nonzero initial conditions imply that a prior, additional injury was applied to the cells. This leads to a very important distinction that will be used in studying the nonautonomous theory. The ***natural system*** is the family of time courses from initial condition (0, 0) and represents the intrinsic ("natural") behavior of the injured system. The ***perturbed systems*** are any time courses from nonzero initial conditions. Our conceptual model of a perturbed system corresponds to preconditioning or hypothermia, as discussed above.

### III. THE NONAUTONOMOUS THEORY OF CELL INJURY

The autonomous model was developed from Eq. (2) by positing that the threshold terms were functions of $I$ as shown in Eq. (3). The nonautonomous version of the cell injury theory is developed by converting the $v$ and $k$ parameters of Eq. (2) to functions as now described.

#### A. Nonautonomous model parameters

Treating parameter $v$ as a constant serves to scale the accumulation rates of $D$ and $S$. However, because we idealize application of injury, $I$, as an instantaneous event at time zero, we imagine some initial rate ($c_{v1}$) that will slow with time after application of the injury, i.e.

$$v_D = v_S = c_{v1} e^{-c_{v2} t} . \tag{6}$$

We stress Eq. (6) is merely a starting point for the sake of study by taking a simple exponential form of time decay. We also assume the rate parameters $v_D$ and $v_S$ are equal for the simplicity of study.

In Eq. (4), $D$ and $S$ are modeled to have first order decay independent of each other. Given that $D$ and $S$ are mutually antagonists, each serving to eliminate the other, the assumption of their independent decay is untenable. The function used to represent the decay parameter, $k$, should couple $D$ and $S$, and derive from considering the physical meaning of the fixed points ($D^*$, $S^*$) in the autonomous model. The bifurcation diagrams (Fig. 1) show that as the system approaches $I_X$ from either side, the magnitude of the difference $|D^* - S^*|$ shrinks, becoming zero at $I_X$. The magnitudes $D^*$ and $S^*$ represent the final amounts of $D$ and $S$ at the completion of the winner-take-all competition between them. There are three cases to consider:
1. $D^* \gg S^*$. Here, the total amount of damage greatly exceeds the total stress responses, thereby overwhelming the cell's ability to cope. It is expected that disintegration of the cell is rapid under such conditions, leading to a large magnitude of decay parameter $k$.
2. $S^* \gg D^*$. Here, the total stress responses greatly exceed the total amount of damage. Since the amount of damage is small compared to the stress responses, we expect rapid clean-up and repair, and fast recovery to the preinjury state, yielding again a large decay parameter.
3. $D^* > S^*$ or $S^* > D^*$ with similar magnitude. If one of $D^*$ or $S^*$ only slightly exceeds the other, as shown in Fig. 2, it takes relatively longer time for their interactions to play out. Thus, the time it takes the cell to die or to recover to the preinjury state will be longer, resulting in a smaller $k$ value.

These considerations lead to the inference that the rate to either death or recovery is a function of $|D^* - S^*|$. We then generalized this consideration by having the decay parameter, $k$, change as a function of the instantaneous value of |D-S| along the entire time course. The new parameter $c_K$ is assumed equal for $D$ and $S$ for simplicity, and thus

$$k_D = k_S = c_k |D - S|. \tag{7}$$

Substituting Eqs. (6) and (7) into Eq. (4) provides the nonautonomous formulation of the cell injury theory:

$$\begin{aligned}
\frac{dD}{dt} &= c_{v1} e^{-c_{v2}t} \frac{(c_D I e^{I\lambda_D})^n}{(c_D I e^{I\lambda_D})^n + S^n} - c_k |D - S| D, \\
\frac{dS}{dt} &= c_{v1} e^{-c_{v2}t} \frac{(c_S I e^{-I\lambda_S})^n}{(c_S I e^{-I\lambda_S})^n + D^n} - c_k |D - S| S.
\end{aligned} \tag{8}$$

**B. Solving the nonautonomous model**

Figure 3 shows example time courses obtained by numerical solution of Eq. (8), where Fig. 3(a) illustrates a survival outcome and 3(b) a death outcome. The time courses begin and end at $(D, S) = (0, 0)$, recapitulating the entire duration from application of injury at time zero to the final outcome $(D^*, S^*) = (0, 0)$ as $t \to \infty$. Eq. (8) overcomes one of the main weakness of Eq. (4) such that every solution to Eq. (8) always ends at $(D^*, S^*) = (0, 0)$, i.e., the biologically meaningful end state as discussed in Sec. II D. Therefore, the standard analysis of fixed points, determination of the separatrix, etc., as was performed for the autonomous model Eq. (4), is inapplicable to the nonautonomous model, given only one trivial fixed point $(0, 0)$ for Eq. (8).

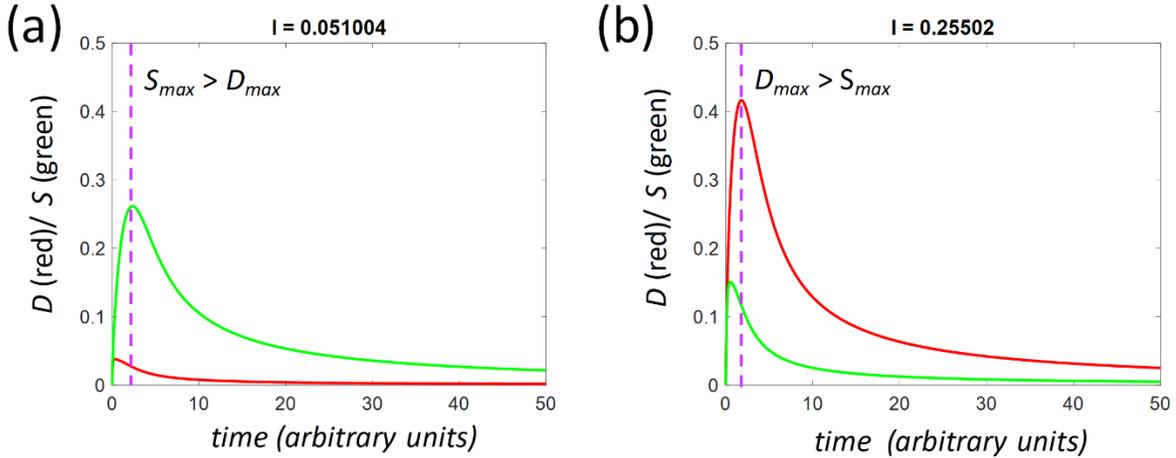

FIG 3. Time courses calculated from the nonautonomous model from initial condition $(0, 0)$. (a) $I < I_X$ and the outcome is survival. (b) $I > I_X$ and outcome is death. Dashed line in (a) marks the location for $S_{max}$, and dashed line in (b) marks $D_{max}$. For this system, $I_X = 0.1275$ and $(c_D, \lambda_D, c_S, \lambda_S, n, c_{V1}, c_{V2}, c_k) = (0.5, 0.75, 0.625, 1, 4, 1, 1, 1)$.

For the autonomous theory, outcome was unambiguously determined by the values of $D$ and $S$ at the fixed points, i.e., $S^* > D^*$ indicating survival and $D^* > S^*$ indicating death. This criterion could not be used for the solutions of Eq. (8) because $(D^*, S^*) = (0, 0)$ for all time courses. On the other hand, it is apparent in Fig. 3(a) that during the evolution process the $S$ time course is greater than the $D$ time course, indicating a survival outcome. For Fig. 3(b), the $D$ time course dominates, indicating a death outcome.

There are two features of the nonautonomous time courses that might serve to establish outcome: (i)

Compare the maxima of each time course ($S_{max}$ vs. $D_{max}$, as indicated in Fig. 3) such that $S_{max} > D_{max}$ indicates survival and $D_{max} > S_{max}$ indicates death. (ii) Determine the late time course behavior. Since Eq. (8) retains the form of a winner-take-all competition, one of the time courses will eventually dominate, and its value will be greater than the other time course at the late time stage, as can be also seen in Fig. 3. Both methods were tested (see Appendix B) and it was established that determining the late time course behavior is the correct method for assigning outcome. For this reason, the study of Eq. (8) focuses on the transient dynamics, specifically, the late time behavior, but not the final system state.

In addition, we have identified a method to express the solutions to Eq. (8) so that they functionally resemble the bifurcation diagram solutions of Eq. (4) [47]. At each $I$, time courses are calculated across ranges of initial conditions ($0 \leq D_0 \leq D_{0,max}$ and $0 \leq S_0 \leq S_{0,max}$) [see Fig. 4(a)]. Next, an "outcome plane" is constructed by plotting a green or red point, indicating a survival or death outcome respectively, at each initial condition [Fig. 4(b)]. Then, outcome planes are calculated across the $I$-range of interest [Fig. 4(c)]. Finally, the percentage of death outcomes on each outcome plane is plotted vs. $I$ to give a "percent death plot" [Fig. 4(d)]. The outcome planes serve a role analogous to the phase planes of the autonomous model by illustrating how the system behaves across initial conditions at a given $I$. The percent death plot functions analogously to a bifurcation diagram of the autonomous model (e.g., Fig. 1). Like a bifurcation diagram, it allows quick assessment of the system as a function of $I$ while also accounting for initial conditions.

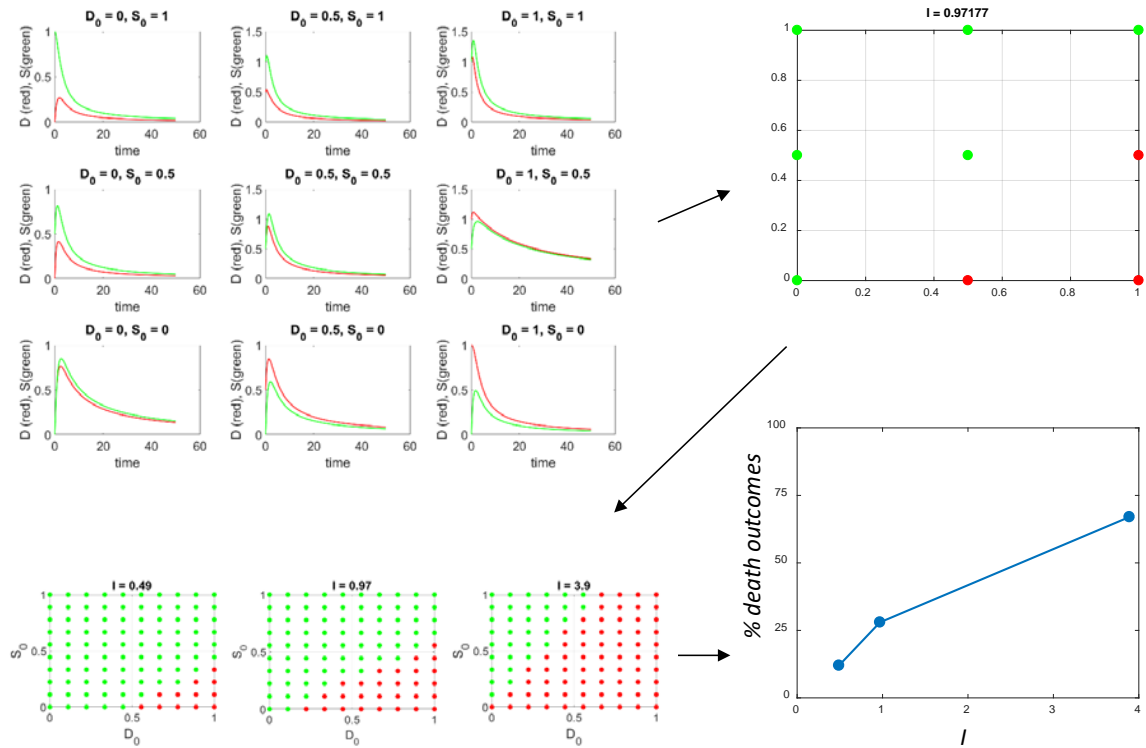

FIG 4. Example of procedure for solving Eq. (8). In (a) $D$ and $S$ time courses are red and green, respectively. In (b) and (c), red and green points represent death and survival outcomes, respectively. The percent death plot (in d) serves as an analog to a bifurcation diagram (e.g., that in Fig. 1). Input parameters for this example are ($c_D$, $\lambda_D$, $c_S$, $\lambda_S$, $n$, $c_{v1}$, $c_{v2}$, $c_k$) = (0.5, 0.75, 15, 1, 4, 1, 1, 1), with $I$ indicated in the plots and the initial conditions spanning the range [0,1].

The novel finding of solving Eq. (8) in this manner is to recognize that, in general, at each $I$, both death and survival outcomes occur. This situation is analogous to bistability in that both death and survival outcomes are present. However, unlike the autonomous model, values of injury intensity containing both outcomes are generally present across the whole $I$-range up to $I_{max}$, after which the outcome is 100% death.

This result overcomes the second limitation of Eq. (4) discussed above in Sec. II D and shows that the dynamics across the *I*-range are a function of both *I* and initial conditions. As seen from the example in Fig. 4(d), the number of death outcomes increases with *I*, as is intuitively expected.

In general, different input parameters to Eq. (8) give qualitatively different percent death plots. Thus, the major goal here is to describe the qualitative dynamics of Eq. (8) across ranges of the input parameters. The details of all the calculations and parameter sweeps are provided in Appendix A, and the main results are described below.

### IV. THE GLOBAL DYNAMICS OF THE NONAUTONOMOUS CELL INJURY THEORY

Figure 5 summarizes the global dynamics of the nonautonomous Eq. (8). We discuss two aspects of the global dynamics: the form of the percent death curves, and the family of time courses starting from initial condition (0, 0), i.e., the family of natural time courses. An important result from studying Eq. (8) is the quantifying of cellular stress responses in the form of the natural maximum total stress response, $S_{max,\text{NAT}}$, and the perturbed maximum total stress response, $S_{max,\text{PER}}$. The definitions of $S_{max,\text{NAT}}$, and $S_{max,\text{PER}}$ are given below while their relevance to stress response biology and protective therapy will be discussed in Sec. V.

#### A. Percent death curves

A total of 5,832 percent death curves were obtained from solving Eq. (8) with 34,992,000 combinations of parameter values (see Appendix A). Studying the form of these revealed only four basic, qualitatively different types of percent death curves. We gave these the descriptive names "hill", "hook", "peak", and "plateau", examples of which are shown in Figs. 5(a)-5(d) (first column of panels). For all four patterns, $I_{max}$ is defined as the lowest value of *I* beyond which 100% death outcomes are achieved. No $I < I_{max}$ exhibited 100% death in any of the percent death curves calculated.

The *hill pattern* [Fig. 5(a)] is the only one of the four to have close to 0% cell death at $I_{min}$. It is called "hill" because it shows a monotonic increase in death outcomes as *I* values increase, with some intervening constant regions, before reaching 100% death outcomes at $I_{max}$. The *hook pattern* [Fig. 5(b)] begins at ~50% death outcome at $I_{min}$, then declines to ~10% death outcome with the increase of *I*, before turning to a monotonic increase with *I* to reach 100% death, thus resembling a hook. The *peak pattern* [Fig. 5(c)] also begins at ~50% death outcome at $I_{min}$, and then increases to reach a "peak" position at ~90% death outcome [arrow, panel 1, Fig. 5(c)], followed by a decline with *I* before turning upwards to 100% death at $I_{max}$. The *plateau pattern* [Fig 5(d)] starts at ~50% death outcome and monotonically increases towards the 100% death outcome at $I_{max}$.

For the hook pattern, there is a range of $I < I_X$ where the percentage of cell death decreases with increasing *I* [orange line in Fig 5(b), panel 1]. For the peak pattern, there is a range of $I > I_X$ where the percentage of death outcomes also decreases with increasing *I* [orange line, Fig 5(c), panel 1]. These are emergent, counter-intuitive results since it is intuitively expected that the cell death percentage will increase with *I*. That only four distinct qualitative patterns were observed stands as a prediction of the theory that any injury to a real cell type will display only one of the four dynamical patterns observed. This will be elaborated in Sec. V.

#### B. Natural Time Courses

The natural time courses for the parameters of the hill, hook, peak, and plateau patterns are plotted across their respective *I*-ranges in Column 2 of Fig. 5. Because these time courses are calculated from initial condition (0, 0), the survival outcome occurs at $I < I_X$ and the death outcome occurs at $I > I_X$.

The natural time courses for the hill pattern shows long durations through a range of about $I = I_X$ to $3I_X$ [indicated by a blue line in Fig. 5(a), panel 2]. Time courses outside this range have shorter durations. Thus, the distinction between slow (delayed death) and rapid (necrosis) death durations is maintained by the nonautonomous model. The hook pattern [Fig. 4(b), panel 2] shows more rapid natural time courses than the hill pattern, but also exhibits variable durations which increase as $I$ approaches $I_X$ from either side. Similarly, the natural time courses for peak and plateau patterns [Figs. 5(c) and 5(d), panel 2] show relatively rapid time courses across the $I$-range, with increased durations around $I_X$.

The hill and hook patterns extended for ~ 6 and 3 $I_X$ units, respectively. The peak and plateau dynamics spanned larger $I$-ranges of ~ 30 and 60 $I_X$ units, respectively. This pattern of relative sizes of $I$-ranges generally held across all percent death curves evaluated. The biological interpretation of this result will be discussed in Sec. V.

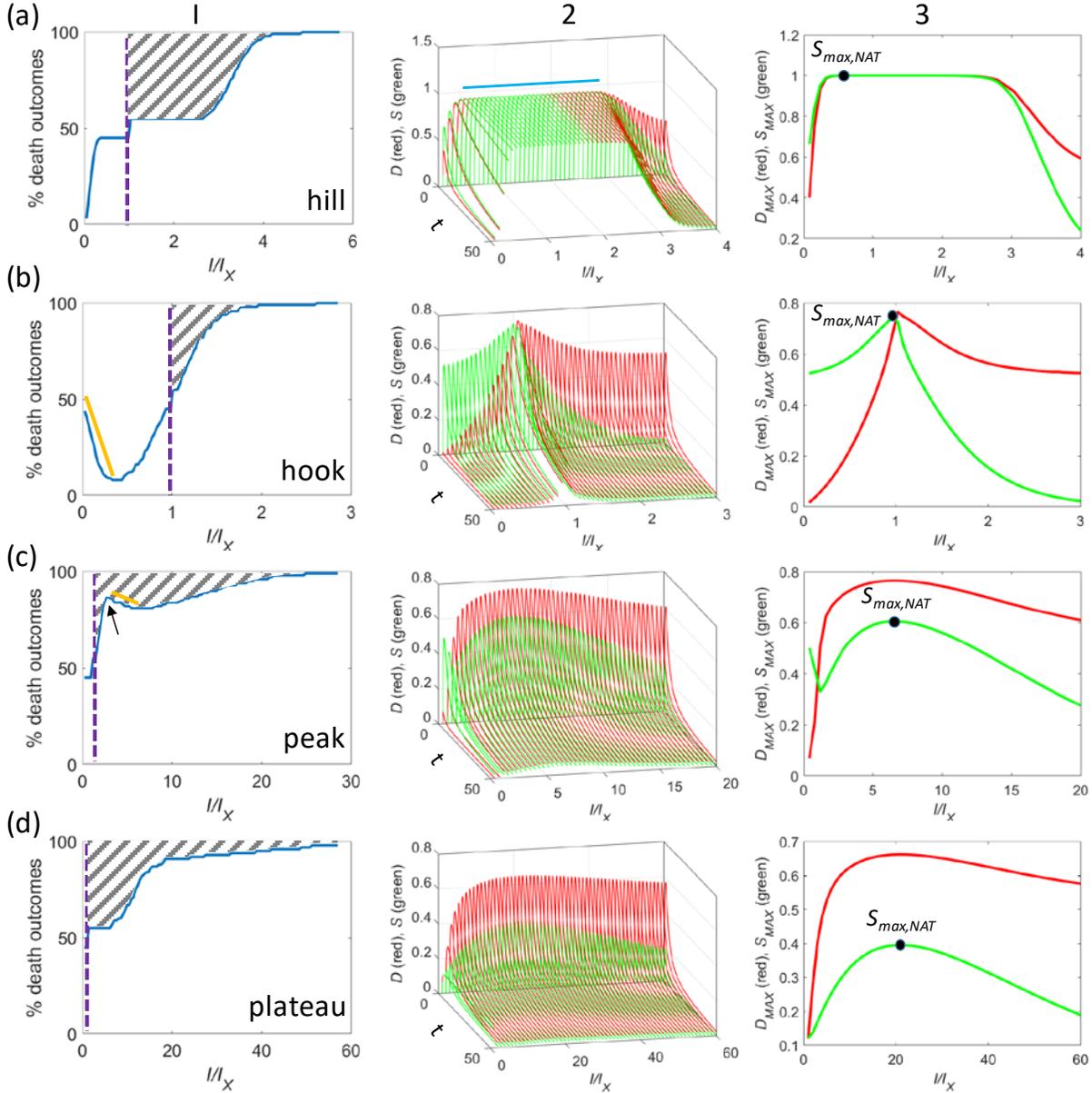

FIG 5. Column 1: Four types of qualitative dynamics obtained from solutions of Eq. (8). $I_X$ is marked by purple dashed line in column 1 panels. The thick orange lines are added to highlight the regions of decreasing percent death. Cross hatched areas mark the ranges of "therapeutic solutions" (see Sec. V B). Column 2: The corresponding natural time courses obtained from the initial condition $(D_0, S_0) = (0, 0)$ (red: $D$ time courses; green: $S$ time courses). Column 3: Plots of the $D$ (red) and $S$ (green) maxima of each time course of column 2 vs $I/I_X$. $S_{max,NAT}$ (black points on green curves) is the largest $S_{max}$ across the $I$-range for the natural time courses. The input parameter vector $(c_D, \lambda_D, c_S, \lambda_S)$ for each row is: (a) (0.5, 0.25, 37.5, 0.25); (b) (0.1, 0.5, 3.5, 1); (c) (0.075, 0.5, 0.1875, 0.1); (d) (0.075, 0.075, 0.0825, 0.075). The percent death curves (column 1) were calculated with $D_0$ and $S_0$ initial conditions over the range [0, 1.5].

### C. Natural Maximum Total Stress Response, $S_{max,NAT}$

As shown in Fig. 3, for a given solution to Eq. (8), both the $D$ and $S$ time courses exhibit a maximum value, $D_{max}$, and $S_{max}$, respectively. The third column in Fig. 5 plots the $I$ dependence of $D_{max}$ and $S_{max}$ obtained from the natural time courses in column 2. It shows that there is one value of $I$ that exhibits a largest value for $S_{max}$, which is termed the *natural maximum total stress response $S_{max,NAT}$*. For the hill and hook dynamics, $S_{max,NAT}$ occurs before $I_X$, while for the peak and plateau dynamics it occurs after $I_X$. As seen in Fig. 1, no such quantity was observed in the autonomous model.

This result indicates that the nonautonomous model generates, and therefore predicts the existence of a quantitative maximum stress response, $S_{max,NAT}$, that would be observable in a given cell type subjected to a specific injury mechanism. Although it is possible to conceive this idea from a qualitative biological perspective, it would lack any framework to ground the notion. To our knowledge, the nonautonomous model, Eq. (8), is the first example of a deductive theory predicting the existence of a maximum total stress response, $S_{max,NAT}$. Furthermore, as elaborated below, the theory also predicts stress response capacity in excess of $S_{max,NAT}$. To develop this idea we need to consider the role of initial conditions in solving Eq. (8).

### D. Outcome as a Function of Initial Conditions

The example of constructing a percent death plot in Fig. 4 shows initial conditions ranging over the interval [0,1]. The question however arises: what is the correct range of initial conditions? Clearly, physical considerations rule out negative values of damage or stress responses, while their upper bounds are unknown. We have numerically tested ranges of initial conditions in the construction of percent death plots and found that: (i) The form of the percent death plot changes as a function of the initial condition range, and (ii) when the initial condition ranges exceed [0, 1.5], the percent death plot lost physical meaning.

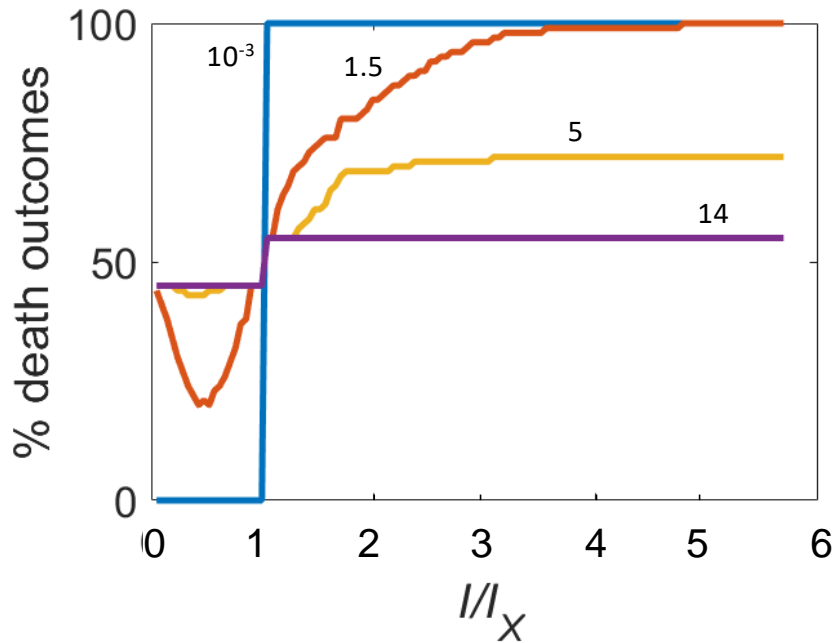

FIG 6. Effect of initial condition ranges on the percent death plot. The primary parameter vector ($c_D$, $\lambda_D$, $c_S$, $\lambda_S$) is chosen as (0.25, 1, 3.125, 1). The number next to each curve is the upper limit of the $D$ and $S$ initial conditions used to construct the corresponding plot.

When the initial condition range approaches zero, the percent death curve is binary: all outcomes at $I <$

$I_X$ give 0% death (i.e., 100% survival) and at $I > I_X$ have 100% death. The limit as the initial condition ranges go to 0 gives, expectedly, the outcome of the natural family of time courses. As the range of initial conditions increases, the percent death curve changes form, through the four qualitative types described above (i.e., hill → hook → peak → plateau). However, in every case tested, when the upper limit of initial conditions exceeded 1.5, the percent death curves never achieved 100% death, no matter how far the *I*-range was extended, a scenario that is clearly unphysical. As seen in Fig. 6, when the upper limit of initial conditions is 5 (orange curve), percent death levels out at ~70% death for all *I*. When the upper limit of initial conditions is set to 14 (purple curve), a binary curve is again obtained where for $I < I_X$ the percent death is 45% and for $I > I_X$ the percent death is 55%. Thus the percent death curve transitions from physically-meaningful binary outcomes of 0%/100% death when the initial condition range approaches zero, to an unphysical binary curve of 45%/55% death when the range of initial conditions exceed 1.5. Mathematically, the value of 1.5 is an arbitrary point depending on the model parameter values chosen. But from the biological viewpoint of cell injury, this value of 1.5 marks the upper limit of physically-applicable initial condition ranges for Eq. (8).

Therefore, we come to a surprising conclusion that the nonautonomous theory dictates the physically-meaningful range of initial conditions. This result held for all the input parameter vectors tested (as given in Appendix A). Thus in this study, including the calculation of percent death curves shown in Fig. 5, the initial conditions were always varied over the range [0, 1.5].

### E. Perturbed Maximum Total Stress Response, $S_{max,\text{PER}}$

Above we established that nonzero initial conditions can be physically interpreted as applying some second form of injury to the natural system, with preconditioning as the canonical example. Preconditioning, hypothermia, drug administration, or any other intervention are perturbations of the natural time courses. Any such perturbation is represented by nonzero initial conditions. Therefore, a percent death curve represents all possible perturbations to the natural system. It was shown above that the natural family of time courses exhibits a maximum stress response $S_{max,NAT}$ under zero initial condition. We asked if there were values of *S* in any of the time course solutions represented by a percent death plot that were greater than $S_{max,NAT}$. The answer was affirmative.

For the 5,832 parameter combinations tested in the study of Eq. (8) (see Appendix A), the value of $S_{max,NAT}$ ranged from 0.148 to 1. For the parameter set giving $S_{max,NAT} = 0.148$, the maximum *S* across all time courses in the initial condition range [0, 1.5] was 1.64. Thus, there exists an *S* time course from nonzero initial conditions that exhibited an 11-fold increase over $S_{max,NAT}$ for that parameter set. For the parameter set giving $S_{max,NAT} = 1$, the maximum *S* across all time courses in the initial condition range [0, 1.5] was 2.5, representing a 2.5-fold increase over $S_{max,NAT}$. Thus, for all the parameter combinations tested, increases of 2.5- to 11-fold over $S_{max,NAT}$ were observed.

These examples illustrate that values of total stress response *S* greater than the natural maximum quantity $S_{max,NAT}$ exist when nonzero initial conditions are considered. The largest of such *S* values we term *the perturbed maximum total stress response*, $S_{max,PER}$. The difference between $S_{max,PER}$ and $S_{max,NAT}$ can be interpreted as a *latent stress response capacity* (LSRC) in injured cells that are subjected to a perturbation, i.e.,

$$LSRC = S_{\max,PER} - S_{\max,NAT}. \tag{9}$$

This has important implications for the mechanism of protective therapeutics as will be discussed ahead.

### V. DISCUSSION

Section I reviewed the empirical findings on acute cell injury. The main conclusion was that the qualitative biomedical approach cannot explain how it is possible to take an injured cell that is fated to die and convert it to a survival outcome, as occurs with preconditioning or hypothermia. It is the purpose of this paper to present the global dynamics of the nonautonomous model that can provide an explanation and corresponding mechanisms. To this end, we showed that the nonautonomous model outputs four types of qualitatively distinct percent death curves, each representing unique patterns of dynamics. The form of the percent death curves depends on the range of initial conditions, where the initial condition range [0, 1.5] sets the limits for a physically-meaningful interpretation of Eq. (8). By studying how the variable $S$ behaves across the initial conditions, $S_{max,NAT}$ and $S_{max,PER}$ were identified. Their difference represents a latent stress response capacity that can be evoked from injured cells by an external perturbation. These results provide additional new insights about the dynamics of acute cell injury, with important implications for eventually applying the theory to assist in developing real-life therapies for medical conditions, such as stroke and heart attack, that have up to now evaded successful treatment.

### A. Phenomenological Models

Throughout we have indicated our simplifying assumptions, idealizations, and choices made for the model development. The alternative, as discussed in Sec. I, are long lists of biological facts that have proven ineffective in the efforts to develop clinical therapies for major acute injuries. The coarse-grained phenomenological models presented here seek to isolate the basic principles behind cell injury and deductively model acute injury in a general way. The ability to retrodict well-established phenomena of cell injury suggests that there is veracity to the principles underlying the phenomenological models. The autonomous model captures the variations in the duration to death after injury, while the nonautonomous model improves on this by calculating dynamic time courses that capture the full cycle of cell injury. The time courses have a relatively simple interpretation: injury displaces a cell out of its steady state. The overall effect of this displacement is characterized by the fixed points in the autonomous model and the values of $D$ and $S$ at late time stage in the nonautonomous model. If $S$ dominates, the cell "falls back" to the preinjury state, but if $D$ dominates the cell "escapes" its own homeostasis and disintegrates (i.e., dies). This is the generic principle captured by the theory that occurs in any cell type injured by any injury mechanism. The additional findings reported here expand on this theoretical conception of cell injury.

### B. The $I$-Range and Therapeutics

A significant difference between the nonautonomous and autonomous models is that both survival and death outcomes are present across the $I$-range in the former but confined to bistable solutions in the latter. For the natural time courses, $I < I_X$ means the cell will survive the injury. But in the nonautonomous model, for $I < I_X$ with $D_0 > 0$, the model outputs death outcomes, as would be expected if there was enough preexisting damage in the cell at the moment of applying the injury. Similarly, at $I > I_X$, the model outputs survival solutions when $S_0 > 0$. Thus, the nonautonomous model gives a more sensible biological result than the autonomous model.

The result also has clinical consequences. With the autonomous model, only the bistable region could be subject to protective therapeutics. As seen in Fig. 1(b), the bistable region is only a fraction of the $I$-range. In contrast, there are survival outcomes across the whole range $I_{min} < I < I_{max}$ of the percent death plots for the nonautonomous model. This provides more opportunity for protective therapeutics across the $I$-range.

For a natural system, $I > I_X$ means the system will die. This case corresponds to an injury that is not subject to therapeutic intervention. However, the nonautonomous model now gives survival solutions over $I_X < I < I_{max}$. We can label survival outcomes at $I > I_X$ as "therapeutic solutions" since they fit the definition of a protective therapy that will halt cell death that, left otherwise untreated, would die. Therapeutic solutions are indicated in the percent death plots in Fig. 5 by the cross hatched areas. As seen, the number of therapeutic solutions decreases as $I$ increases from $I_X$ to $I_{max}$. Thus, it is expected it will be increasingly difficult to access the therapeutic solutions as $I$ approaches $I_{max}$, and likely to be a case of diminishing returns to attempt to extract every last survival outcome.

Survival outcomes at $I < I_X$ are not therapeutic solutions because the natural system survives any $I < I_X$. However, the nonautonomous model indicates that, if the system is perturbed, it is possible for a system at $I < I_X$ to have a death outcome when $D_0 > 0$. This also has clinical relevance in the context of co-morbidities where the presence of a pre-existing disease or injury could cause an additional otherwise nonlethal injury to be lethal. A concrete example of such a case is stroke outcome of diabetics compared to nondiabetics: diabetics who have a stoke suffer greater brain damage than if a similar magnitude of stroke occurs in a nondiabetic [48].

Thus, the nonautonomous model provides quantitative insight into protective therapeutics and comorbidities. Such considerations are natural interpretations of the theory solutions. They are practical, quantitative considerations for the development of protective therapeutics that are inconceivable from within the qualitative biomedical mindset.

### C. Four Qualitative Dynamical Patterns

The four distinct forms for the percent death curves (see Fig. 5, Column 1 panels) can be used to explain how different cell types respond to injury. Cardiomyocytes die in minutes to hours after ischemia [49], whereas, as explained above, neurons die hours, days, and weeks after exposure to equivalent ischemia intensity [26,28]. Additionally, it was explained how preconditioning is carried out differently in brain and heart. Brain preconditioning is called "delayed preconditioning" because of the long durations involved, whereas cardiac preconditioning is called "rapid". Such temporal differences can be linked to the four qualitative types of percent death curves. The hill dynamics show long times to cell death around $I_X$ which is reminiscent of the behavior of neurons after stroke and of neuron preconditioning kinetics. The other three dynamical patterns, hook, peak, and plateau show rapid decay to death when $I > I_X$ and one of these may model the behavior of heart cells after ischemia and how they are preconditioned. In general, the four dynamical patterns provide quantitative templates for the interpretation of experimental data and can also guide the design of data acquisition.

### D. Initial conditions

The form of percent death curves changes with initial condition ranges. In the limit where the initial condition ranges approach zero, all percent death curves reduce to the intuitive and idealized result that 100% survival occurs at $I < I_X$ and 100% death occurs at $I > I_X$ (see Fig. 6 blue curve). This exactly recapitulates the monostable case in the autonomous model [Fig. 1(a)]. On the other hand, in the limit where initial condition ranges go to infinity, we obtained the unphysical result of 45%/55% percent death curves (Fig. 6, purple curve). Unphysical percent death appeared when values of initial conditions exceed 1.5. We therefore conclude that the nonautonomous model self-determines the appropriate range of initial conditions based on the physical interpretation of the outcome. We note also that an initial condition range maximum of 1.5 is only 50% greater the largest values of $D$ and $S$ across the natural time courses. This puts the initial condition range in the same order of magnitude as the time course solutions. More future studies across a wider parameter space would be needed to further understand this result.

### E. Latent Stress Response Capacity: Basic and therapeutic considerations

Currently, cell stress responses are understood mainly in terms of specific qualitative molecular pathways [50]. This level of understanding merely describes and catalogs stress responses. There is a relatively smaller literature that seeks to quantify stress response pathways by treating them with the formalisms of classical chemical equilibria and chemical kinetics [51-53]. This latter approach is limited because: (1) cells are not at equilibrium and it is questionable if equilibrium expressions are applicable, especially following acute cell injury where the system is in a constant state of transformation, and (2) there is no guarantee that all possible molecular interactions are accounted for in the choices of products and reactants. Our theories use the coarse-graining of $D$ and $S$ [e.g., Eq. (1)] to approximate and account for all possible interactions. Further, our equations are phenomenological and agnostic to the status of chemical equilibrium. Thus, while our intention has been to quantify acute cell injury dynamics, the nonautonomous model, via the concepts of $S_{max,NAT}$ and $S_{max,PER}$, appears to provide a foundation for examining the quantitative biology of cellular stress responses.

The existence of a LSRC, Eq. (9), also has consequences for therapeutic technology by providing an explanation of how protective therapeutics is possible. In the autonomous model, protective therapeutics was explained by bistability at $I > I_X$ indicating the system possesses both survival and death attractors (Fig. 1(b)). For the nonautonomous model with its trivial single fixed point (0, 0), this explanation of bistability is not applicable because attractor states no longer factor into the solutions. However, the tradeoff is the existence of the LSRC in the nonautonomous model. This new result suggests that protective therapies such as preconditioning and hypothermia function by accessing the LSRC. The additional amount of $S$ this provides serves to "push" the cell from a pro-death to a pro-survival trajectory.

It bears mention that $S_{max,PER}$ is the highest value of $S$ achievable for a given percent death plot and represents the maximum possible total stress response the cell can exert. However, $S_{max,PER}$ cannot be accessed in the natural state [i.e., the family of time courses from initial condition (0, 0)]; it can only be accessed via a second manipulation, such as preconditioning or hypothermia. This means that, operationally, $S_{max,PER}$ cannot be experimentally measured in the natural state and measuring it requires some additional perturbation of the cells other than the injury mechanism, such as by preconditioning or hypothermia.

### F. Future Directions for Theory Development

The present analysis of the nonautonomous model provides a building block for constructing progressively more realistic models of cell injury, with the ultimate goal to construct models of specific forms of injury to specific organs. Eq. (8) can be used to construct "multiple injury models" (MIMs) that simulate applying different injuries over time as occurs with preconditioning and hypothermia. We have shown initial results of a MIM construction elsewhere [46]. The result obtained here, that there is an intrinsic limit to the range of initial conditions, will affect MIM constructions. How to integrate this result into MIM construction is an important future direction of study. An additional future direction is the construction of models with an explicit spatial dependence. We envision spatial models where the cellular distribution of an organ will be approximated by a three-dimensional distribution of grid points, where each grid point models a single cell. The injury will then be modeled by a spatial and temporal distribution of injury intensity, $I$, superimposed over the grid of cells. An initial discussion of spatial models of cell injury has been provided elsewhere [46]. The understanding of the nonautonomous model described here moves us closer to constructing spatial models that will allow applying the theory to multi-cellular structures.

### VI. CONCLUSIONS

We have examined the global dynamics of a nonautonomous formulation of the acute cell injury theory, which is a phenomenological theory intended to capture the basic principles governing how a living cell responds to acute injury. The theory recapitulates important salient aspects of the empirical facts and predicts the existence of four distinct qualitative dynamics and of a latent stress response capacity of cells. As a nonlinear dynamical theory, it is unique for its focus on the transient dynamics that determines the outcomes and mechanisms of the cell injury.

This deductive approach has been undertaken because the currently dominant qualitative mind-set in biomedical research has not been effective at achieving its stated goal of developing protective therapies to halt cell death in injured tissues. The development and study of the cell injury theory is a first step towards its practical applications, particularly the development of quantitative and systematic medical therapeutic technologies. The issues involved in operationalizing the theory are daunting, but not insurmountable. A future report will detail our attempts to operationalize the model through experimental measurements and analyses.

**Acknowledgments**

This work was supported by the National Institutes of Health, National Institute of Neurological Disorders and Stroke, under Grant No. NIH R21NS081347 (D.J.D.) and the National Science Foundation under Grant No. DMR-1609625 (Z.-F.H.).

**Appendix A: Details of Numerical Calculation**

Equation (8) cannot be solved analytically and thus numerical solutions were undertaken in Matlab using ode45 and the built-in parallel computing functions. (The corresponding Matlab code can be obtained by request to the first author (DeGracia) via email.) The workflow was designed to optimizing computational time, minimize floating point errors, and maximize the range of parameters tested. The general workflow was the following: A total of 34,992,000 different values of input parameter vector ($c_D$, $\lambda_D$, $c_S$, $\lambda_S$, $n$, $c_{V1}$, $c_{V2}$, $c_K$, $I$, $D_0$, $S_0$) were used to generate $D$ and $S$ time courses and the corresponding outcomes. Percent death plots were constructed across various $I$-ranges. Because there were 27 $c_S$ chosen per $c_D$ (see below), continuation plots of 27 percent death curves for each of the 216 combinations of ($c_D$, $\lambda_D$, $\lambda_S$) were used to organize the time courses. A typical continuation plot is illustrated in Fig. 7. The set of 216 continuation plots was repeated seven times across different ranges of initial conditions. The specifics are as follows.

**1. Initial Parameter Vector 1**

The 216 combinations of ($c_D$, $\lambda_D$, $\lambda_S$) were constructed from:

$c_D$ = [0.075  0.1  0.25  0.5  0.75  1]
$\lambda_D$ = [0.075  0.1  0.25  0.5  0.75  1]
$\lambda_S$ = [0.075  0.1  0.25  0.5  0.75  1]

By Eq. (5), a strict parameter constraint of the model is $c_S > c_D$ so that $I_X > 0$. Thus, parameter $c_S$ was chosen as multiples of $c_D$. Our studies showed that: (1) percent death plots tended to become uniform after $c_S \approx 75 c_D$ (taking on the hill dynamic), and (2) the following 27-valued choice of $c_S$ led to reasonably uniform spacing on log($c_S$) plots:

$c_S = c_D$[1.1, 1.25, 1.5, 1.75, 2, 2.5, 3, 3.5, 4, 5, 6, 8, 10, 12.5, 15, 20, 25, 30, 35, 40, 45, 50, 55, 60, 65, 70, 75]

In addition, the following parameters were held constant for all parameter sets:

$n = 4$; $c_{V1} = 1$; $c_{V2} = 1$; $c_K = 1$

where $c_{v1}$ and $c_K$ scale the time course heights. Setting them to 1 constrains $D$ and $S$ to the unit plane. Parameters $n$ and $c_{V2}$ were kept constant to simplify the analysis. Thus, initial parameter vector 1, IPV1, is defined as ($c_D$, $\lambda_D$, $c_S$, $\lambda_S$, $n$, $c_{V1}$, $c_{V2}$, $c_K$).

**2. Determination of *I*-ranges**

The *I*-range ($I_{min} \leq I \leq I_{max}$) was determined for each IPV1 because $I_{max}$ is specific for a given IPV1, recalling that $I_{max}$ is the value of $I$ after which there is only 100% death outcome for all subsequent $I$. To be computationally efficient, a two-step procedure was used to determine $I_{max}$.

Step 1 incremented $I$ in units of $I_X$ [as calculated by Eq. (5)] and calculated the $D$ and $S$ time courses only from the single initial condition $(D_0, S_0) = (0, 0)$. $I$ was incremented until the $S$ time course was effectively zero across all the time. The value of $I$ at which this occurred was $I_{test}$.

Step 2 calculated the outcome plane at $I_{test}$ across a range of 100 initial conditions (as specified in the next subsection). If the outcome plane was 100% death, $I_{test}$ was decremented by 1 $I_X$ unit until the outcome plane was <100% death and the prior value of $I$ was taken as $I_{max}$. If the outcome plane at $I_{test}$ was <100% then $I_{test}$ was incremented in one $I_X$ unit, taking as $I_{max}$ the first $I$ to give an outcome plane of 100% death. Thus, $I_{max}$ was determined to within one $I_X$ unit.

Once $I_{max}$ was estimated, the $I$-range was divided into 60 equal increments (e.g. $I_{min} = I_{max}/60$), and each IPV1 was spawned into 60 initial parameter vector 2, IPV2, across $I_{min} \leq I \leq I_{max}$: ($c_D$, $\lambda_D$, $c_S$, $\lambda_S$, $n$, $c_{V1}$, $c_{V2}$, $c_K$, $I$).

### 3. Initial Conditions

Seven initial condition ranges, $0 \leq D_0 \leq D_{0,max}$ and $0 \leq S_0 \leq S_{0,max}$, with $D_{0,max} = S_{0,max}$, were studied over the interval [0, 2]. Each initial condition range was divided into 10 equal increments, and a 10×10 grid of all combinations were constructed, giving 100 initial conditions per IPV2. Negative initial condition values were not studied.

### 4. Summary of parameter choices

Eq. (8) was solved 34,992,000 times using input vector ($c_D$, $\lambda_D$, $c_S$, $\lambda_S$, $n$, $c_{V1}$, $c_{V2}$, $c_K$, $I$, $D_0$, $S_0$). The quantitative summary of the choices of parameter vectors is:

(1) 216 combinations of ($c_D$, $\lambda_D$, $\lambda_S$)
(2) 27 $c_S$ per ($c_D$, $\lambda_D$, $\lambda_S$) combination $\rightarrow$ 216 × 27 = 5,832 combinations
(3) 60 values across $I$-range $\rightarrow$ 5,832 × 60 = 349,920
(4) 100 initial conditions ($D_0$, $S_0$) per ($c_D$, $\lambda_D$, $\lambda_S$, $c_S$, $I$) $\rightarrow$ 100 × 349,920 = 34,992,000
(5) $n = 4$; $c_{V1} = 1$; $c_{V2} = 1$; $c_K = 1$ for each run of the 34,992,000 parameter vectors.

Most results reported in this paper were from the set of 34,992,000 runs with initial condition maxima $D_{0,max} = S_{0,max} = 1.5$. Additional calculations have been conducted for other 6 sets (each having these 34,992,000 parameter combinations) when choosing different values of $D_{0,max} = S_{0,max}$, with some results summarized in Table 1 of Appendix B.

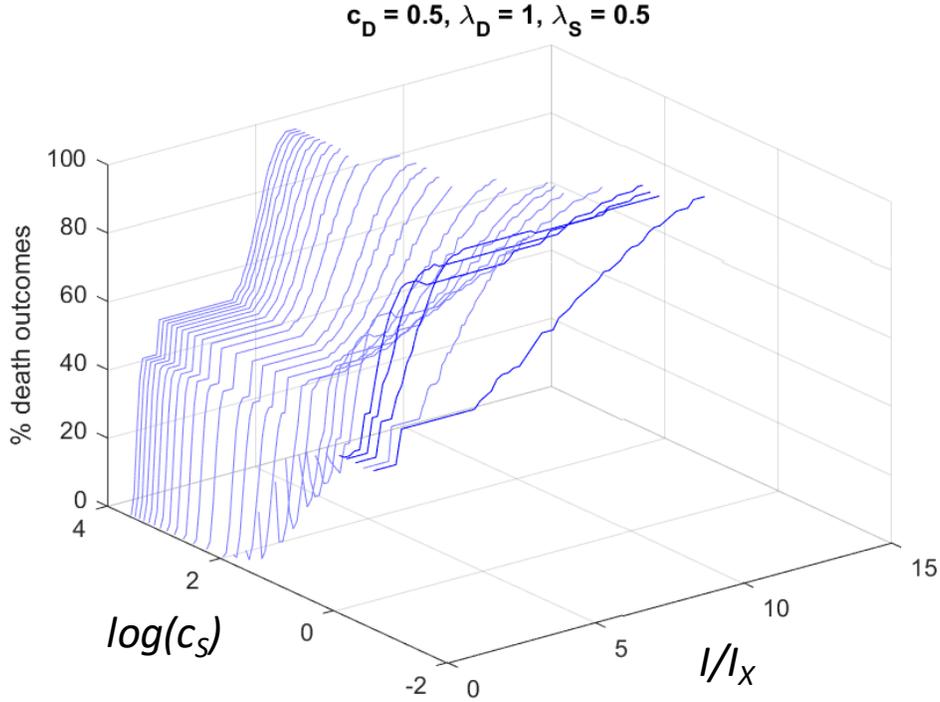

**FIG 7**: Example continuation plot of a family of 27 percent death plots, used to identify the qualitative dynamics of Eq. (8). The main parameters $c_D$, $\lambda_D$, and $\lambda_S$ are indicated, and values of $c_S$ were set as multiples of $c_D$ as explained in section A.1. Each percent death curve is calculated to its respective $I_{max}$. The $I$-range is expressed as $I/I_X$. $D_0$ and $S_0$ initial conditions were over the range [0, 1.5]. The percent death plots form a continuous transition from the plateau → peak → hook → hill dynamic as $c_S$ increases. This result held for all 216 continuation plots which varied in: (1) the length of the $I$-ranges, and (2) the extent of the $c_S$ range occupied by each of the four qualitative dynamics.

### 5. Time Course Durations

Given around two orders of magnitude differences in the chosen range of several parameters, each input vector had unique durations for the $D$ and $S$ time courses. For every run of Eq. (8) described above, the time course solutions were passed through a subroutine to determine the ratios $D_{end}/D_{max}$ and $S_{end}/S_{max}$, where $D_{end}$ and $S_{end}$ were the last points computed for each time course (representing the results at late time stage), and $D_{max}$ and $S_{max}$ were the maximum values of the respective time course. These ratios had to be < 1% to accept a pair of time courses as "completed". If they failed this test, the time course duration was incremented by a factor of 5 until this test was passed. The minimum time course range was 50 time units in 0.1 increments, and the maximum was 3,906,250 time units in 7812.5 increments.

### Appendix B: Outcome Determination

Each of the 34,992,000 parameter vectors used in Eq. (8) produced a pair of $D$ and $S$ time courses. For Eq. (8), at $t = \infty$, $(D, S) = (0, 0)$. However, at late time stage with large $t$ (e.g., at the end of each time course numerically calculated, with results $D_{end}$ and $S_{end}$), the dominant time course is always greater than the subdominant, which served as our effective definition of a winner-take-all model. Therefore, if $D_{end} > S_{end}$, the outcome was death, and if $S_{end} > D_{end}$, the outcome was survival. Given the above criteria for determining time course durations (Appendix A.5), $D_{end}$ and $S_{end}$ were well within floating point limits.

In our previous work presenting the method to express the solutions to Eq. (8) [47], we used the maximum points of the $D$ and $S$ time courses to determine the cell outcome where $D_{max} > S_{max}$ corresponded to death and $S_{max} > D_{max}$ to survival. However, in the present study the extensive exploration of initial conditions revealed that maxima comparisons were inadequate across the parameter ranges. As shown in Fig. 8, when values of initial conditions were greater than or equal to time course maxima for $t > 0$, the overall maximum point would be the initial one at $t = 0$. When this occurred, it might or might not agree with the determination of outcome using the late-stage points of the calculated time courses.

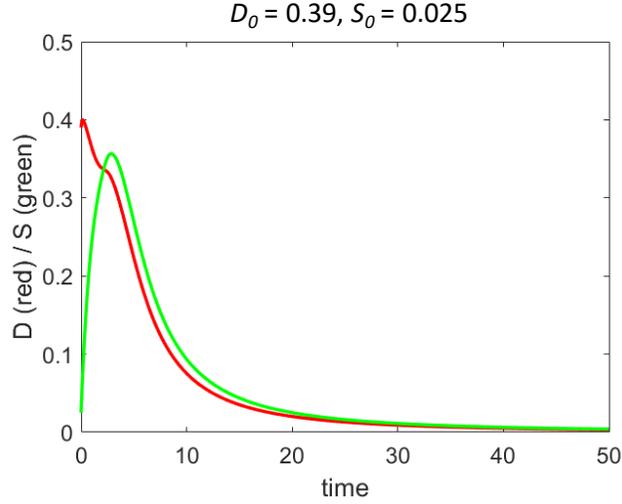

**FIG 8**: Example time course in which the initial condition, $D_0$, was greater than any point along the $D$ time course. In this example, at late time stage the $S$ time course (green curve) is seen to dominate and therefore this solution to Eq. (6) is taken as a survival outcome. Input parameter vector ($c_D$, $\lambda_D$, $c_S$, $\lambda_S$, $I$) is (0.075, 0.25, 1.5, 0.75, 0.28).

We quantified this disagreement on outcome determination and found, expectedly, the number of disagreements decreased as the range of initial conditions decreased, as shown in Table 1, column 3, where the initial condition range is expressed in terms of the maximum value of $S$ across the family of time courses calculated from initial conditions (0, 0), $S_{max,NAT}$ (column 1). Further, the effect was ubiquitous across primary parameter vectors ($c_D$, $\lambda_D$, $c_S$, $\lambda_S$), affecting over 95% of them when the maximum initial condition was $> 0.25 S_{max,NAT}$ (column 2). Thus, in the present study we used the late-stage results of time course (i.e., end points of the numerical calculation) to determine outcome and did not use time course maxima.

**TABLE 1**: Quantification of outcome mismatches as a function of initial condition ranges. In Column 1, initial conditions are expressed as fold of $S_{max,NAT}$. Column 2 shows the percentage of the 5,832 ($c_D$, $\lambda_D$, $c_S$, $\lambda_S$) primary parameter vectors (PPV) displaying mismatches in outcome determination between the maxima and late-stage methods (a PPV is counted as containing mismatch as long as any one of the ($I$, $D_0$, $S_0$) combinations corresponding to it displayed the mismatch). Column 3 lists the percentage of all the ~35 million input parameter vectors displaying outcome mismatches.

| $S_{0,max}/S_{max,NAT}$ | % of PPV | % mismatches |
|---|---|---|
| 2.00 | 99.9% | 5.08% |
| 1.00 | 99.8% | 2.62% |
| 0.50 | 99.2% | 0.85% |
| 0.25 | 96.4% | 0.33% |
| 0.125 | 86.7% | 0.15% |
| 0.05 | 76.2% | 0.06% |